\documentclass[traditabstract]{aa}

\usepackage{graphicx}
\usepackage{txfonts}
\usepackage{natbib}
\usepackage[colorlinks=true,citecolor=blue]{hyperref}
\usepackage{xspace}
\usepackage{array}
\usepackage[labelformat=simple]{caption,subcaption}
\usepackage{bm}
\usepackage{amsmath}
\usepackage{txfonts}
\usepackage{ulem}
\usepackage{multirow}
\usepackage{soul}

\bibpunct{(}{)}{;}{a}{}{,}

\newcommand{\mic}{$\muup$m\xspace}

\newcommand{\lsd}{\hbox{$\lambda_0/D_{pup}$}\xspace}
\newcommand{\up}[1]{\textsuperscript{#1}}

\usepackage{color}

\begin{document} 

\title{Low-order wavefront control using a Zernike sensor \\ through Lyot coronagraphs for exoplanet imaging}
\subtitle{Blind stabilization of an image dark hole}

\author{
    R.~Pourcelot\inst{\ref{oca}}     
    \and
    M.~N'Diaye\inst{\ref{oca}}
    \and
    E.~H.~Por \inst{\ref{stsci}}
    \and
    I.~Laginja \inst{\ref{lam}, \ref{onera}}
    \and
    M.~Carbillet\inst{\ref{oca}}  
    \and
    H.~Benard \inst{\ref{tas}}
    \and
    G.~Brady \inst{\ref{stsci}}
    \and
    L.~Canas \inst{\ref{tas}}
    \and \\
    K.~Dohlen \inst{\ref{lam}}
    \and 
    J.~Fowler \inst{\ref{stsci}}
    \and
    O.~Lai \inst{\ref{oca}}
    \and 
    M.~Maclay \inst{\ref{stsci}}
    \and
    E.~McChesney \inst{\ref{stsci}}
    \and 
    J.~Noss \inst{\ref{stsci}}
    \and
    M.~D.~Perrin \inst{\ref{stsci}}
    \and
    P.~Petrone \inst{\ref{hf}}
    \and \\
    L.~Pueyo \inst{\ref{stsci}}
    \and 
    S.~F.~Redmond \inst{\ref{Princeton}}
    \and
    A.~Sahoo \inst{\ref{stsci}}
    \and
    A.~Vigan \inst{\ref{lam}}
    \and
    S.~D.~Will \inst{ \ref{stsci}, \ref{Rochester}, \ref{gsfc}}
    \and
    R.~Soummer \inst{\ref{stsci}}
}

\institute{
    Université Côte d’Azur, Observatoire de la Côte d’Azur, CNRS, Laboratoire Lagrange, France \\
    \email{\href{mailto:raphael.pourcelot@oca.eu}{raphael.pourcelot@oca.eu}} \label{oca}
    \and
    Space Telescope Science Institute, 3700 San Martin Drive, Baltimore, MD 21218, USA \label{stsci}
    \and
    Aix Marseille Université, CNRS, CNES, LAM (Laboratoire d’Astrophysique de Marseille) UMR 7326, 13388 Marseille, France
    \label{lam}
    \and
    ONERA, The French Aerospace Lab, BP72, 29 avenue de la Division Leclerc, 92322 Ch\^atillon Cedex, France \label{onera}
    \and
    Thales Alenia Space, 5 Allée des Gabians - B.P. 99 - 06156 Cannes la Bocca Cedex – France \label{tas}
    \and
    Hexagon Federal, Chantilly, VA 20151, USA \label{hf}
    \and 
    Department of Mechanical and Aerospace Engineering, Princeton University, Princeton, NJ 08540, USA \label{Princeton}
    \and
    The Institute of Optics, University of Rochester, Rochester, NY 14627, USA \label{Rochester}
    \and
    NASA Goddard Space Flight Center, Greenbelt, MD 20771, USA \label{gsfc}
    }

\date{\today}

\abstract
    {Combining large segmented space telescopes, coronagraphy and wavefront control methods is a promising solution to produce a dark hole (DH) region in the coronagraphic image of an observed star and study planetary companions. The thermal and mechanical evolution of such a high-contrast facility leads to wavefront drifts that degrade the DH contrast during the observing time, thus limiting the ability to retrieve planetary signals.}
    {Lyot-style coronagraphs are starlight suppression systems that remove the central part of the image for an unresolved observed star, the point spread function, with an opaque focal plane mask (FPM). When implemented with a flat mirror containing an etched pinhole, the mask rejects part of the starlight through the pinhole which can be used to retrieve information about low-order aberrations.} 
    {We propose an active control scheme using a Zernike wavefront sensor (ZWFS) to analyze the light rejected by the FPM, control low-order aberrations, and stabilize the DH contrast. The concept formalism is first presented before characterizing the sensor behavior in simulations and in laboratory. We then perform experimental tests to validate a wavefront control loop using a ZWFS on the HiCAT testbed.}
    {By controlling the first 11 Zernike modes, we show a decrease in wavefront error standard deviation by a factor of up to 9 between open- and closed-loop operations using the ZWFS. In the presence of wavefront perturbations, we show the ability of this control loop to stabilize a DH contrast around $7\times10^{-8}$ with a standard deviation of $7\times10^{-9}$.}
    {Active control with a ZWFS proves a promising solution in Lyot coronagraphs with an FPM-filtered beam to control and stabilize low-order wavefront aberrations and DH contrast for exoplanet imaging with future space missions.}


\maketitle

\section{Introduction}
With about 5000 known planets orbiting nearby stars\footnote{http://exoplanet.eu/}, finding an Earth analog around a Sun-like star has seen an increasing interest for the past few years. Still, to assess the resemblance of a given exoplanet candidate with Earth, characteristics such as atmospheric composition and physical parameters need to be retrieved. Exoplanet direct imaging and spectroscopy represent a promising approach to collect photons from a planetary companion around its parent star and perform its spectral characterization. However, this proves extremely challenging as exoplanets, and in particular Earth twins, are up to 10\up{10} times fainter than their host star at angular separations shorter than 100\,mas in the visible wavelength range. 

Since the mid 2010s, circumstellar disks and planetary companions have been routinely observed around nearby stars with several state-of-the-art exoplanet imagers on 8-m class ground-based telescopes \citep{Macintosh2014, Jovanovic2015, Beuzit2019} and their large surveys \citep[e.g.,][]{Nielsen2019,Vigan2021}. They have so far enabled the observation of young gas giant planets with a planet-to-star flux ratio, or contrast, down to 10\up{-6} at angular separations as small as 200\,mas from their host star in the near infrared \citep[e.g.,][]{Macintosh2015, Chauvin2017, Keppler2018, Currie2020}. The next generation of large ground-based facilities includes the Extremely Large Telescopes \citep[ELTs,][]{Simard2016, Fanson2018, Ramsay2020} to further understand the formation and evolution of planetary systems. Equipped with instruments to image small gaseous or large telluric planets and spectrally analyze their atmosphere, they are expected to achieve contrasts down to 10\up{-8} at distances shorter than 50\,mas in the near infrared \citep{Fitzgerald2019, Guyon2019, Kasper2021}. In addition, future large observatories in space with high-contrast capabilities are envisioned options to observe small rocky planets with 10\up{-10} contrast at less than 50\,mas in reflected light in the visible.

Following this path, in its 2020 Astronomy and Astrophysics Decadal Survey, the United States National Academies of Science, Engineering, and Medicine (NASEM) recommends the development of an at least 6-m diameter space telescope, large enough to provide the sensitivity and resolution for the detection and characterization of Earth-like exoplanets \citep{2020DecadalSurvey}. Such a large telescope is expected to be deployable and have a segmented primary mirror to fit current and expected launch vehicles. This mission will build on the recent studies from two NASA mission concepts, the Habitable Exoplanet Imaging Mission \citep[HabEx,][]{Habex2019} and the Large Ultraviolet Optical Infrared surveyor \citep[LUVOIR,][]{LUVOIR2019}. While HabEx has an approach based on a 4-m diameter monolithic-mirror telescope, LUVOIR relies on a concept with a segmented primary mirror with a diameter of up to 15\,m.

To overcome the large flux ratio between an observed star and its planet, several teams are currently investigating different schemes of high-contrast instruments with coronagraphic capabilities. Preliminary coronagraph designs, simulations, and experiments show 10\up{-9} to 10\up{-10} raw contrast capabilities, with the assumption of very low wavefront errors or sufficient photon flux for wavefront correction loops \citep[e.g.,][]{Ruane2018b, N'Diaye2016, Por2020, Sirbu2021, Seo2019}. Carried out to refine the stability specifications for LUVOIR, the ULTRA study presents a total wavefront error budget of 158\,pm Root Mean Square (RMS) at all timescales and for aberrations with all spatial frequency contents \citep{Ultra2019}. To meet these constraints, active optics with Deformable Mirrors (DMs) prove a promising solution for two main reasons: (i) to produce a high-contrast region, or dark hole (DH), in the coronagraphic image of an observed star to enhance the planet signal, (ii) to maintain this DH during observations by correcting for the aberrations induced by thermal and mechanical evolution of the observatory and measured by wavefront sensors (WFS) in real time. 

The first-step active optics system leads to the generation of a DH with an optimal wavefront shape to reach the targeted contrast. However, this wavefront shape experiences small perturbations due to mechanical flexures and thermal variations during the observing time. These perturbations lead to aberrations that degrade the achieved DH contrast. Depending on the coronagraph design, several second-step wavefront sensing schemes and control solutions have been proposed in recent years to stabilize the DH contrast during the observation, \citep[e.g.,][]{Jovanovic2018}. Among these wavefront sensing solutions, the Zernike wavefront sensor \citep[ZWFS, e.g.,][]{Zernike1934, Bloemhof2003, Dohlen2004, Wallace2011, N'Diaye2013} has proven to be one of the most efficient sensors in terms of photon usage \citep{Guyon2005, Chambouleyron2021b}, capable of measuring aberrations down to the picometer level \citep{Ruane2020, Steeves2020}. In the context of the Nancy Grace Roman Space Telescope \citep[RST,][]{Spergel2015,Krist2015}, this solution was adopted in the presence of a monolithic primary mirror to control line-of-sight, jitter pointing errors, thermal focus drifts and other low-order modes. On RST, the ZWFS uses the light rejected in reflection by the focal plane mask (FPM) of the Hybrid Lyot Coronagraph or the Shaped Pupil Coronagraph \citep{Shi2016, Shi2018}.

In the case of segmented aperture telescopes, such as LUVOIR, recent developments have considered Lyot-type Coronagraphs (LCs) as promising solutions for starlight suppression. LCs use a FPM that can be implemented using a flat mirror with a pinhole to reject starlight. Such a configuration has been studied in the Lyot project \citep{Sivaramakrishnan2007} and was successfully implemented on the exoplanet imagers Palomar P1640 \citep{Oppenheimer2012} and the Gemini Planet Imager for exoplanet observations \citep{Macintosh2014}. Taking advantage of the available photons from the FPM-filtered beam, we explore the use of a second-step active control loop using a ZWFS to control low-order aberrations. The aim is to maintain the wavefront shape at a point that yields a minimum DH contrast.

In this paper, we present the concept of a ZWFS in a LC FPM-filtered beam, the formalism to describe the optical propagation, and the analytical phase reconstruction of the low-order aberrations with our ZWFS-based approach. Using the High-contrast imager for Complex Aperture Telescopes testbed \citep[HiCAT,][]{hicat1, hicat2, hicat3, hicat4, hicat5, hicat6}, we implement and test our approach for experimental validations in the presence of a segmented aperture. The ZWFS response is studied for the first 11 Zernike modes and with different DM offsets, which correspond to the commands that are sent to the DMs to either maximize the optical wavefront quality or produce an image DH on HiCAT. Finally, we show the stability results for the wavefront error and DH contrast that are achieved with our concept, in the presence of drifts introduced on DMs or with various configurations of internal air turbulence on the bench. Specific measurements of segment piston, tip and tilt errors are not considered in this study and left for a future analysis.

\section{Zernike wavefront sensor for a FPM-filtered beam}
\begin{figure*}
    \centering
    \includegraphics[width=\linewidth]{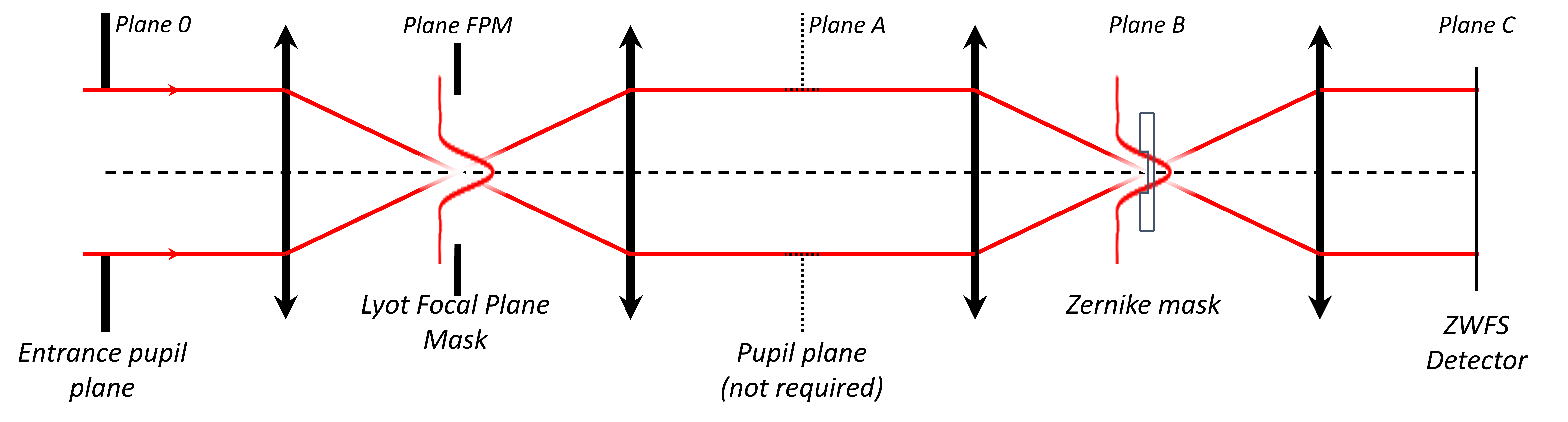}
    \caption{Schematic diagram of the ZWFS downstream of a binary amplitude FPM. The light in the entrance pupil comes from the telescope and deformable mirrors in plane 0. It is focused on the FPM, that is re-imaged on the Zernike mask in plane B. The intermediate pupil plane A is not required.}
    \label{fig:sch_diagram}
\end{figure*}

In the context of LCs, we explore strategies to use the light rejected by their FPM with a ZWFS to control low-order aberrations. Figure~\ref{fig:sch_diagram} shows a schematic diagram of this configuration. The complex amplitude of the electric field $\psi_0$ in the pupil plane 0 at the entrance pupil of the system is spatially filtered by the FPM in the downstream focal plane: we study the light that goes through the mask to the re-imaged pupil plane A. From this plane onward, the propagation is identical to the standard ZWFS described by \citet{N'Diaye2013}. The filtering step induced by the FPM modifies the electric field, and this impact will be detailed in the following section. We then present the corresponding analytical phase reconstruction and the closed-loop strategy to control low-order wavefront errors.

\subsection{Formalism of the ZWFS in a filtered beam} \label{sec:formalism}

We present the formalism that relates the electric fields in the entrance pupil and on the ZWFS camera. The formalism is detailed in two parts: first by describing the interference induced by the Zernike phase mask, and then by detailing the effect of the FPM filtering.

Derived from the phase contrast technique proposed by \citet{Zernike1934}, the ZWFS uses a phase-shifting mask that is located in a focal plane in which the point spread function (PSF) of an unresolved source is formed. Its typical size is one resolution element $\sim \lambda/D$, where $\lambda$ and $D$ denote the source wavelength and the aperture diameter, respectively. This mask, usually manufactured as a glass plate with a small phase dimple, produces interference between the central core of the PSF that goes through the dimple and the rest of the light that passes outside the dimple. The light propagation through the ZWFS can be described using the formalism presented in \citet{N'Diaye2013}: we call $\psi_A$ the complex amplitude of the electric field in the relay pupil plane A between the FPM and the Zernike phase mask. This complex amplitude can be decomposed into a pupil amplitude $P$ and phase $\varphi$ such that
\begin{equation}
    \psi_A = P\,e^{i\varphi}.
\end{equation}

In the downstream focal plane B in which the Zernike mask is inserted, the complex amplitude $\psi_B$ can be expressed as
\begin{equation}
    \psi_B = t\,\widehat{\psi}_A,
\end{equation}
where $\widehat{f}$ denotes the Fourier transform of $f$, and $t$ the transmission of the Zernike mask. If $M_Z$ is the top-hat function describing the mask dimple with a value of 0 outside the mask and 1 inside, and $\theta$ the phase shift introduced by the mask, $t$ can be expressed as
\begin{equation}
    t = 1 - \left(1-e^{i\theta}\right)M_Z.
\end{equation}

For the standard ZWFS, $\theta$ is usually equal to $\pi$/2 but the formalism remains valid for any value that is not a multiple of $2\pi$ (absence of a phase mask shift).
Finally, the complex amplitude $\psi_C$ on the ZWFS camera in the re-imaged pupil plane C can be written as
\begin{align} 
    \psi_C &= \widehat{\psi}_B \\
           &= \psi_A - \left(1-e^{i\theta}\right) \widehat{M}_Z\otimes\psi_A  \label{eq:zwfs_general} \\
           &= \psi_A - \left(1-e^{i\theta}\right)b\,, 
\end{align}

with $\otimes$ the convolution product and $b = \widehat{M}_Z\otimes \psi_A$, the wave diffracted by the Zernike mask. The intensity $I_C = |\psi_C|^2$ on the detector can then be expressed as a function of $\varphi$:
\begin{equation} \label{eq:final_equation}
    I_C = P^2 + 2b^2(1-\cos\theta) + 2Pb\left[\sin\varphi\sin\theta-\cos\varphi(1-\cos\theta)\right].
\end{equation}

This expression does not make any assumption on $P$: the geometry and the amplitude of the pupil can be arbitrary and therefore our formalism is also valid for a pupil containing a gray-scale apodization. Filtering the beam by a FPM does not change the relation between the phase in the intermediate pupil A and the intensity on the detector.

We define $\varphi_0$ as the phase of the wave in the entrance pupil plane 0 and $\psi_0$ as its electric field:
\begin{equation} \label{eq:psi_0}
    \psi_0 = P_0 \, e^{i\varphi_0}\,.
\end{equation}
The next step is to find a relation between the phase error $\varphi_0$ in the entrance pupil plane 0 and $\varphi$, the phase in plane A. In the general case, this link can be established through the electric fields $\psi_0$ and $\psi_A$: 

\begin{equation} \label{eq:psiA_to_psi0}
    \psi_A = \widehat{M}_{\rm{FPM}} \otimes \psi_0,
\end{equation}
where $M_{\rm{FPM}}$ is the top-hat function of diameter $d_{\rm{FPM}}$ that describes the FPM pinhole.
In the general case, the relation between $\varphi_0$ and $\varphi$ cannot be established in a straightforward way and more complex strategies have to be used, such as a data-driven calibration with a Jacobian matrix, as detailed in Sec.~\ref{sec:closed_loop_operations}.

Another aspect of this description deals with the properties of $b$ in the presence of the FPM, which can be implemented in reflection (e.g., a mirror with a pinhole) or in transmission. By replacing $\psi_A$ in the definition of $b$ by Eq.~\ref{eq:psiA_to_psi0}, we obtain:
\begin{align}
    b = \widehat{M}_Z\otimes\widehat{M}_{\rm{FPM}} \otimes \psi_0.
\end{align}
In this equation, $\widehat{M}_Z\otimes\widehat{M}_{\rm{FPM}}$ is the Fourier transform of $M_Z\times M_{\rm{FPM}}$, with $\times$ the element-wise multiplication. For most applications, the Zernike phase dimple diameter $d_Z$ is smaller than the FPM pinhole diameter in angular units: $d_Z \approx 1\,\lambda/D$ and $d_{\rm{FPM}}$ is larger than 4\,$\lambda/D$. In this case, $M_Z\times M_{\rm{FPM}}$ is equal to $M_Z$. As a result $\widehat{M}_Z\otimes\widehat{M}_{\rm{FPM}}$ can be simplified as $\widehat{M}_Z$. The wave diffracted by the mask can therefore be given by
\begin{align}
    b = \widehat{M}_Z \otimes \psi_0.
\end{align}

In the regime of small aberrations ($\varphi_0\ll$\,1\,rad), the central core of the PSF can be considered unaberrated, meaning that $b$ is constant, as described in \citet{N'Diaye2013}. With the expression of $\psi_0$ in Eq.~(\ref{eq:psi_0}), $b$ can be written as
\begin{align}
    b = \widehat{M}_Z \otimes P_0.
\end{align}
Under these simplifying assumptions, the term b hence only depends on the entrance pupil aperture and the Zernike phase mask, but not on the FPM. This aspect proves useful for the reconstruction method detailed in the following section.

\subsection{Analytical reconstruction} \label{sec:analytical_reconstruction}

We now present the inversion of Eq.~(\ref{eq:final_equation}) to retrieve $\varphi$ from the measurement $I_C$. 

In the small-aberrations regime ($\varphi\ll$\,1\,rad), we can use a second-order Taylor expansion of $\varphi$ as detailed in \citet{N'Diaye2013} to derive the following expression:
\begin{equation} \label{eq:2ndorder}
    I_C = P^2 + 2\,b^2(1-\cos{\theta}) + 2\,P\,b\left[\varphi \sin{\theta} - (1-\varphi^2/2)(1-\cos{\theta})\right].
\end{equation}

This quadratic equation is presented as a good approximation in the $[-\pi/4,\,\pi/4]$ range for $\varphi$.

Assuming $P$ and $b$ are non null, this second-degree polynomial equation can be solved by computing the discriminant $\Delta$ with
\begin{align}
    \Delta = \sin^2{\theta} - \frac{2 (b-P) (1-\cos^2{\theta})}{P}  - \frac{P^2-I_C(1-\cos{\theta})}{P b} .
\end{align}
Finally, $\varphi$ can be retrieved using:
\begin{align}
    \varphi = \frac{-\sin{\theta}+\sqrt{\Delta}}{1-\cos{\theta}}.
    \label{eq:phase_second_order}
\end{align}

In the general case (without any assumption on $\varphi$), it is possible to derive the exact expression for the phase $\varphi$ from Eq.~(\ref{eq:final_equation}), yielding
\begin{equation} \label{eq:exact_inversion}
    \varphi = \arcsin{\left[\frac{I_C - P^2 - 2b^2(1-\cos{\theta})}{4Pb\sin{\frac{\theta}{2}}}\right]} + \frac{\theta}{2}.
\end{equation}

By using either the 2nd-order approximation or the exact inversion, the phase reconstruction always requires an estimation of $P$ and $b$. The latter can be computed using a numerical propagation of the entrance pupil plane amplitude $P_0$. To eliminate systematic errors by using an analytical input, $b$ can also be computed by using the pupil amplitude in plane A $P$, as $b$ is not modified by the FPM filtering. A measurement of $|P|^2$ can be recorded on the ZWFS camera by simply moving the Zernike mask out of the beam. In most optical systems, the assumption $P = \sqrt{|P|^2}$ is valid as the electric field in the entrance pupil is positive or null for all kinds of pupils and apodizations. However, in the presence of FPM filtering, the pupil amplitude $P$ is not necessarily positive or null in the pupil plane, since the FPM filtering induces $\pi$-phase shifts in some areas of the pupil measurement. To still make use of the empirical estimation of $P$, an analytical pupil sign mask input has to be provided which allows us to determine the sign of $P$ all over the pupil. Such a mask can be obtained by numerically propagating $P_0$ through the optical system.  

\begin{figure}
    \centering
    \includegraphics[width=\columnwidth]{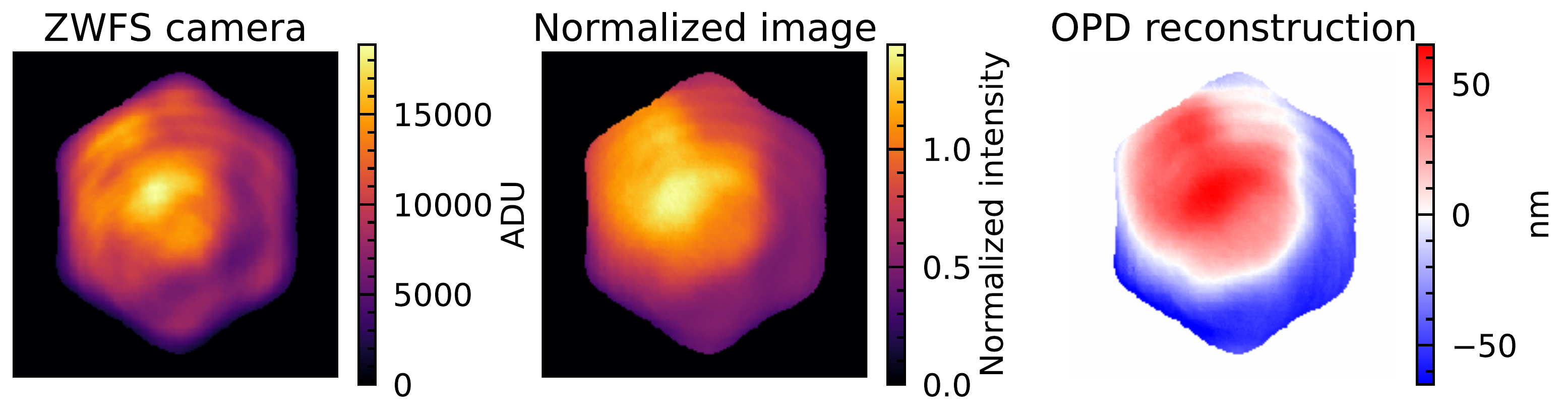}
    \caption{Example of intensity map as measured by the ZWFS (left), normalized by the $|P|^2$ measurement (middle), and the reconstructed phase map using the 2nd-order analytical reconstructor (right).}
    \label{fig:image_examples}
\end{figure}

Fig.~\ref{fig:image_examples} shows an example of the image $I_C$ recorded on the HiCAT bench, normalized by $|P|^2$, and the reconstructed wavefront $\varphi$. Normalizing the image removes some artifacts coming from pupil illumination discrepancies. The full phase reconstruction requires using Eq.~(\ref{eq:phase_second_order}) or Eq.~(\ref{eq:exact_inversion}) including the estimation of $b$. 

Throughout this paper, we use the open-source code pyZELDA \citep{Vigan2018b} with the updated formalism of \citet{N'Diaye2013} presented here to take into account the FPM filtering. Our code performs phase reconstruction using Eq.~(\ref{eq:phase_second_order}), as we are working with small phase errors.

\subsection{Closed-loop sensing and control} \label{sec:closed_loop_operations}

We assume that a DM is located in an upstream pupil plane conjugated with plane 0 to inject Zernike modes in our system, and to control wavefront errors.

The phase reconstruction using our analytical method based on phase conjugation can be used as a wavefront error estimator in closed-loop for wavefront stabilization \citep{Pourcelot2021}. Here we work with the approach based on the use of a Jacobian matrix $\textbf{J}$, often called interaction matrix. This proves suitable for enhanced control-loop stability, in particular for long experiments in which an improper alignment between the Zernike camera and the DM may cause divergence in the phase reconstruction process. The Jacobian matrix approach is a powerful way to relate the measured filtered aberrations to the standardized unfiltered Zernike modes. In addition, its Singular Value Decomposition (SVD) gives the singular modes which can easily be filtered to select the modes to control. The phase reconstruction with our analytical method is used as an additional diagnostic tool, in particular to estimate the total wavefront error (WFE) in a real-life system.  

The data-driven calibration for this Jacobian matrix is performed by poking the DM actuators with $m$ modes from a suitable basis. A DM probe introducing an optical path difference (OPD) value $p$ is used and the corresponding image ${I}_C$ in its vectorized version is recorded within a pre-established region of interest (ROI) with $n$ pixels on the ZWFS camera. For each mode $i$, the intensity responses to a positive and a negative probe are recorded, yielding $P_+^i$ and $P_-^i$ measurements for ${I}_C$. The matrix $\textbf{J}$ is therefore constructed by concatenating the $(P_+^i - P_-^i) / 2p$ vectors for the $m$ modes. As this matrix is rectangular, we perform a SVD to produce its pseudo-inverse $\mathbf{C}$, the control matrix. Using the SVD, the Jacobian matrix is expressed as 

\begin{align}
    \mathbf{J} = \mathbf{U}~\mathbf{S}~\mathbf{V}^T, 
\end{align}
where $\mathbf{S}$ is a diagonal matrix whose values are the singular values of $\mathbf{J}$. The matrices $\mathbf{U}$ and $\mathbf{V}$ are square unitary matrices of respective dimensions $m\times m$ and $n\times n$. The matrix $\mathbf{C}$ is then obtained by computing
\begin{align}
    \mathbf{C} = \mathbf{V}~\mathbf{S}^{+}~\mathbf{U}^{T},
\end{align}
where $\mathbf{S}^{+}$ denotes the pseudo-inverse of $\mathbf{S}$, with $\mathbf{S}^{+}=(\mathbf{S}^{T}~\mathbf{S})^{-1}~\mathbf{S}$.

In this expression, the diagonal matrix $\mathbf{S}^{+}$ contains the inverse coefficients of the non-null $\mathbf{S}$ coefficients. The smallest singular values of $\mathbf{S}$ correspond to poorly sensed modes or noisy artifacts that require large strokes on the DM to be controlled. They can be set to zero without the loss of too much information. The matrix $\mathbf{S}^{+}$ can therefore be replaced by its filtered version $\mathbf{S}_{k}^{+}$ for which $k$ represents the number of the first ordered singular values that are left non-nulled. Correspondingly, $\mathbf{C}$ is replaced by $\mathbf{C}_k$ by substituting $\mathbf{S}^{+}$ with $\mathbf{S}_{k}^{+}$. This filtering helps to stabilize the control loop by avoiding to  introduce correction errors. 

For a given image $I_C$ within the ROI, we obtain the list of coefficients $\hat{\alpha}$ with
\begin{equation}
    \hat{\alpha} = \mathbf{C}_k I_C \,.
\end{equation}
These coefficients $\hat{\alpha}$ are associated with the $k$ modes and allow us to produce a phase map to apply to the DM shape for modal control. 
In a real-life system, a control loop gain smaller than 1 is used to ensure its convergence towards a null wavefront error with the control algorithm. 

\section{ZWFS characterization}
We characterize the behavior of the ZWFS in the presence of a FPM-filtered beam by using the setup that was recently installed on HiCAT. 

\subsection{Experimental setup on HiCAT}

HiCAT is a high-contrast testbed to develop and mature exoplanet imaging technologies for future large segmented-aperture telescopes. Its goal is to demonstrate a complete systemic approach, combining a segmented aperture, coronagraphy, and wavefront sensing and control to achieve a $10^{-8}$ contrast in air and in broadband light in the visible. This testbed is installed in a clean-room environment and inside an enclosure that is stabilized in humidity and temperature by a valve-controlled air flow. Tuning it allows us to change the air turbulence strength. Several light sources are available on HiCAT and for our experiments we work in monochromatic light at the wavelength $\lambda_0=$640\,nm.

\begin{figure*}[!ht]
    \centering
    \includegraphics[width=\linewidth]{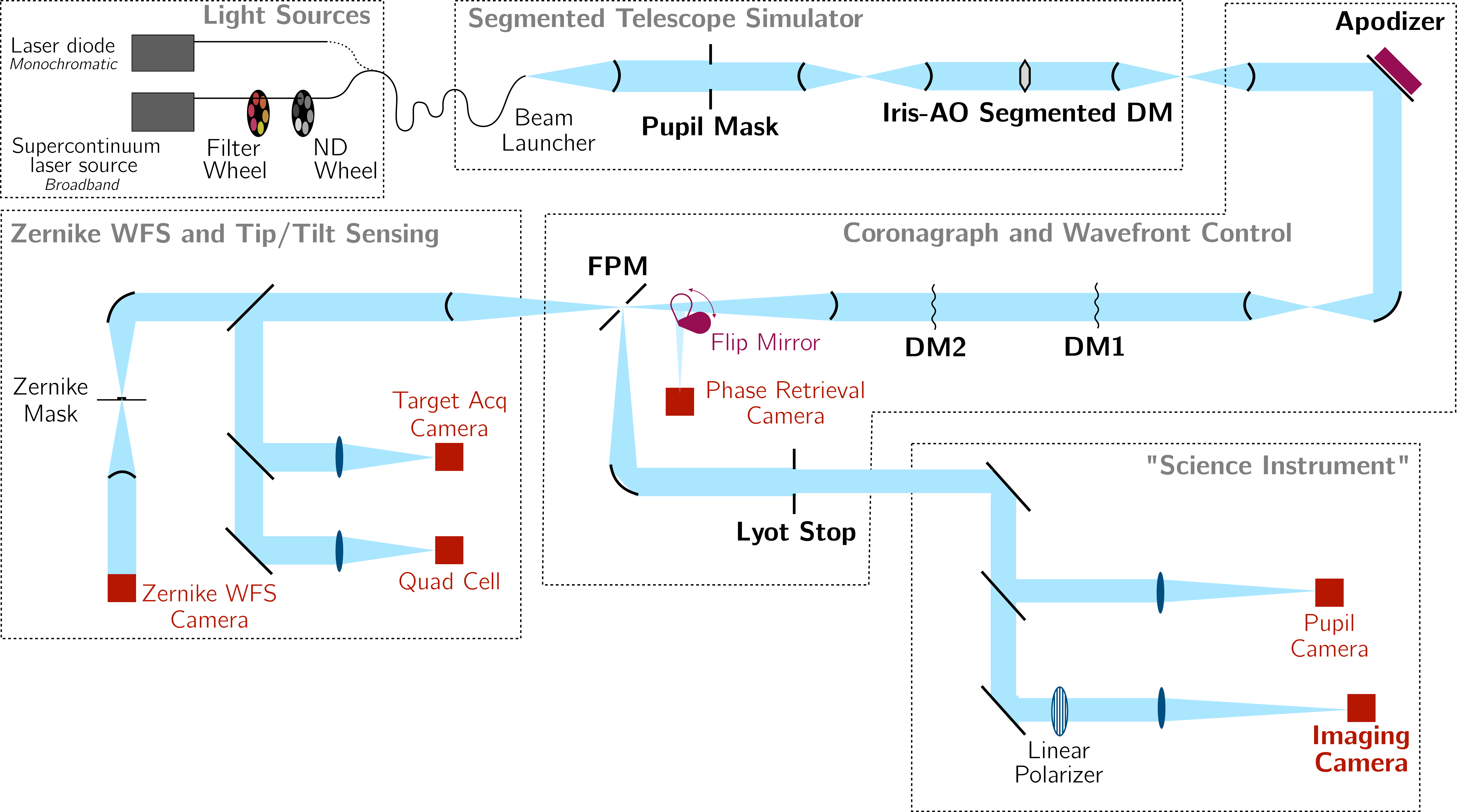}
    \caption{Semi-unfolded layout of the HiCAT testbed. The apodizer labelled in the diagram top right corner is currently replaced by a flat mirror.}
    \label{fig:hicat_blueprint}
\end{figure*}

The current design of the testbed is presented in the partially unfolded diagram in Fig.~\ref{fig:hicat_blueprint}. Among different components, HiCAT includes an IrisAO PTT111L deformable mirror with 37 segments to mimic a segmented aperture. By default, the IrisAO uses a calibrated flat map to minimize the WFE due to segmentation. The current starlight suppression system is a Classical Lyot Coronagraph (CLC) composed of a reflective FPM made up of a flat mirror with a central pinhole, and a Lyot Stop (LS). The imaging camera, a ZWO ASI178, is used to display the coronagraphic image of the source with a sampling of 10.5\,pix per \lsd, with $D_{pup}$ denoting the entrance pupil diameter. A dedicated pupil plane is also available to insert an apodizer and implement the Apodized Pupil Lyot Coronagraph (APLC) \citep[APLC, e.g.][]{Soummer2005, N'Diaye2016}, but in our experiment this component is replaced with a flat mirror. The FPM pinhole diameter is $d_{\rm{FPM}} = 455\,\mu$m, leading to an angular size of $8.52\,\lambda_0/D_{pup}$, with an entrance pupil diameter $D_{pup}$ of 19.55\,mm. The beam transmitted through the FPM pinhole is caught by off-axis parabolas (OAPs) to feed several sensors: a Target Acquisition (TA) camera in a focal plane for FPM centering, a Phase Retrieval camera (PR), and a ZWFS for low-order aberration measurements. 

The ZWFS mask consists of a fused silica substrate with an ion-etched dimple in the center which is aligned with the source image in the associated focal plane. The Zernike dimple has a depth of 280\,nm and a diameter of 54.3\,\mic, leading to an angular diameter $d_{Z}=1.02\,\lambda_0/D_{pup}$. The mask was manufactured by SILIOS Technologies using photo-lithography with a reactive ion-etching process, as described in \citet{N'Diaye2010}. The ZWFS camera is a ZWO ASI178 with $3096 \times 2080$ pixels and a pitch of 2.4\,\mic. The speed limit of the camera is 60\,Hz in full-frame operations. As the source provides enough photons, the typical exposure time is 0.5\,ms in our experiments.

\begin{figure}[!ht]
    \centering
    \includegraphics[width=\columnwidth]{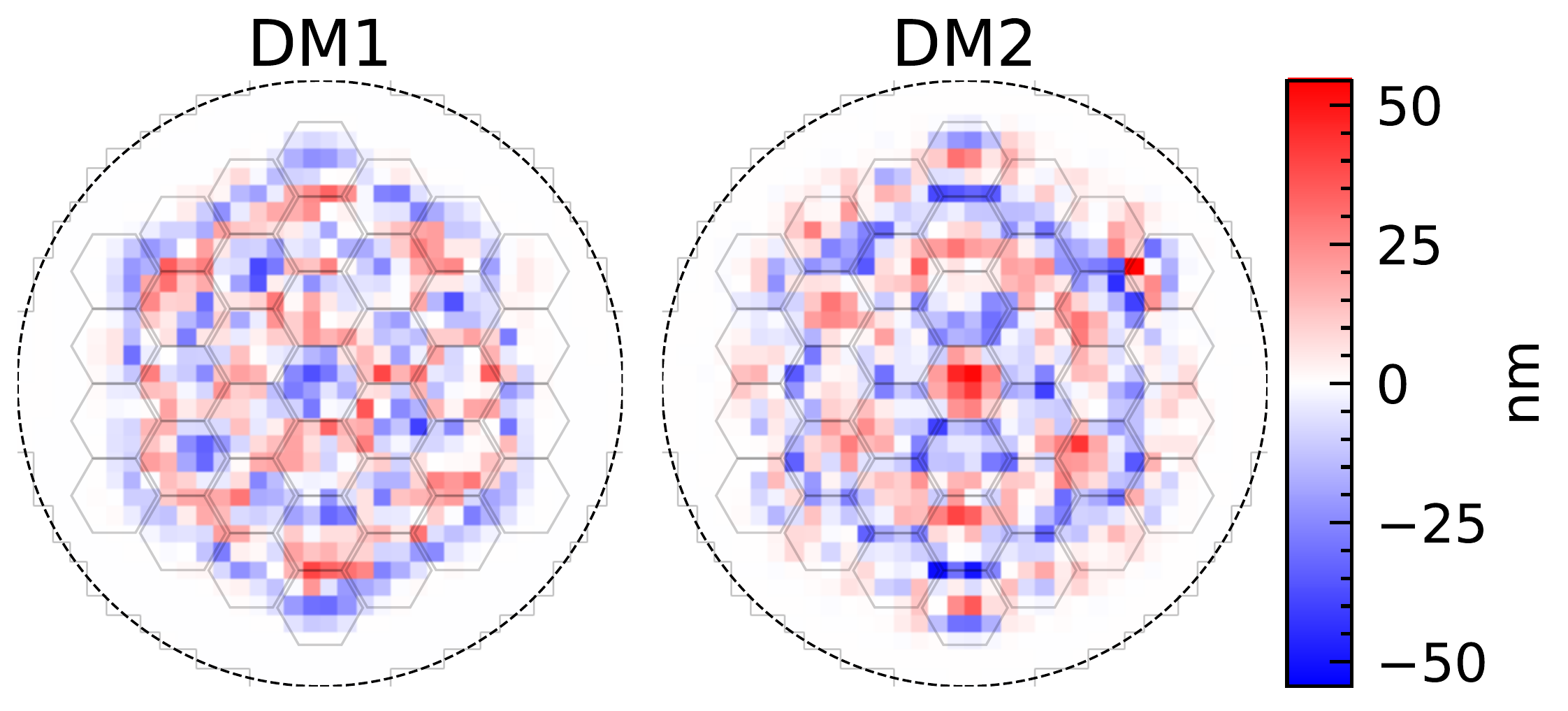}
    \caption{Example of DM commands on both continuous deformable mirrors produced by the SM algorithm that can generate a DH on HiCAT. The DM commands in units of OPD is an approximation of the introduced deflection. The black circle represents the active area of the continuous DMs. For scaling comparison, IrisAO segments are plotted in transparency. DM2 being out of pupil plane, the segment footprints are not realistic here.}
    \label{fig:sm_shapes}
\end{figure}

To shape the wavefront, HiCAT uses two continuous Boston kilo-DMs: DM1 in a pupil plane and DM2 out of a pupil plane to allow both phase and amplitude control. A prior calibration of HiCAT has been performed to experimentally produce DM flat commands, or flat maps, that minimize the amount of static aberrations on the testbed \citep{2018SPIE10698E..6IB}. HiCAT also uses both of these DMs to further reject starlight and enhance the contrast in a given region of the coronagraphic image by using DH optimization algorithms. Optimal DM commands for DH generation are obtained with the stroke minimization (SM) algorithm \citep{Pueyo2009, Mazoyer2018b, Mazoyer2018a}. An example of typical DM commands producing a DH with this algorithm is shown in Fig.~\ref{fig:sm_shapes}, with deflections ranging from approximately -50\,nm to +50\,nm. These commands produce a DH ranging from $4.5$ to $7.5\,\lsd$ in the focal plane, as operated on HiCAT as of December 2021. When these commands were generated, they produced a DH with a $2.5\times 10^{-8}$ average contrast. Our current work aims to reduce the inner working angle (IWA) while keeping the same contrast level but this is out of the scope of this paper. Throughout this paper and as handled on HiCAT, we present DM commands as the OPD that we aim to introduce. The DM commands in volt are computed using the conversion factors from nm to volt that were measured with a Fizeau interferometer. These gains were computed by using the voltage ratio between the flat map and a pure defocus map. These defocus map voltages have been calibrated with the Fizeau interferometer to produce a pure defocus. The expected and introduced OPD will mainly differ because of actuator coupling and non-linearity for large strokes.

For our experiments, we operate our wavefront control loop using the ZWFS with two different DM settings as a starting point: (i) with the DM flat maps providing the optimal optical quality of the testbed and (ii) with the DM maps as provided by SM producing a coronagraphic image DH. Both settings will allow us to monitor the wavefront and DH stabilization with our active control loop. When using DM SM maps, we apply a set-and-forget operation mode: we first run the SM algorithm to obtain the DM commands and then run the ZWFS experiment. The delay between the SM and ZWFS experiments is no longer than a few minutes to prevent the testbed conditions from evolving too much in the meantime. The SM and ZWFS operations have so far been performed separately and successively. Further developments will enable to combine SM control and ZWFS closed-loop control simultaneously.

In the rest of the paper, $\mathbf{J}$ is computed by using $n$ at around 24000 pixels, the exact number depending on the selected ROI. The DM probes that introduce Zernike modes in OPD are applied on top of the flat maps in case (i) or on top of the SM commands in case (ii). From empirical tests to optimize the control, we use $p$=20\,nm RMS, $m=12$ and we only filter out the smallest singular value to keep $k=11$ modes. We use the Noll convention \citep{Noll1976} for Zernike modes, poking the first 12 modes from piston to vertical secondary astigmatism. The DM response to piston, which is not a pure piston, is poorly sensed by the ZWFS. Its sensitivity to a given mode can be computed as the variance of the coefficients of the corresponding column in $J$. In our experiments, the ZWFS sensitivity to piston is 18 times smaller than the averaged sensitivity to the other 11 modes. Similarly, it is five times smaller than the sensitivity to trefoil, which is the second less sensed mode. For better loop stability, this requires filtering at least one mode in the inversion of $\mathbf{J}$. Piston is only kept in the probes within the control code for simplicity in operations. In our experiment, the requested pokes are pure Zernike modes. Their difference with the introduced modes is calibrated by $\mathbf{J}$. Further studies will be required to assess the maximum number of Zernike modes to be controlled or to address alternative modes that may be more suitable for this control. Finally for the control loop, the implemented controller is an integrator, with an integral gain of 1 and a proportional gain of 0.15. 

\subsection{Modal response curves}
\begin{figure*}[!ht]
    \centering
    \includegraphics[width=\linewidth]{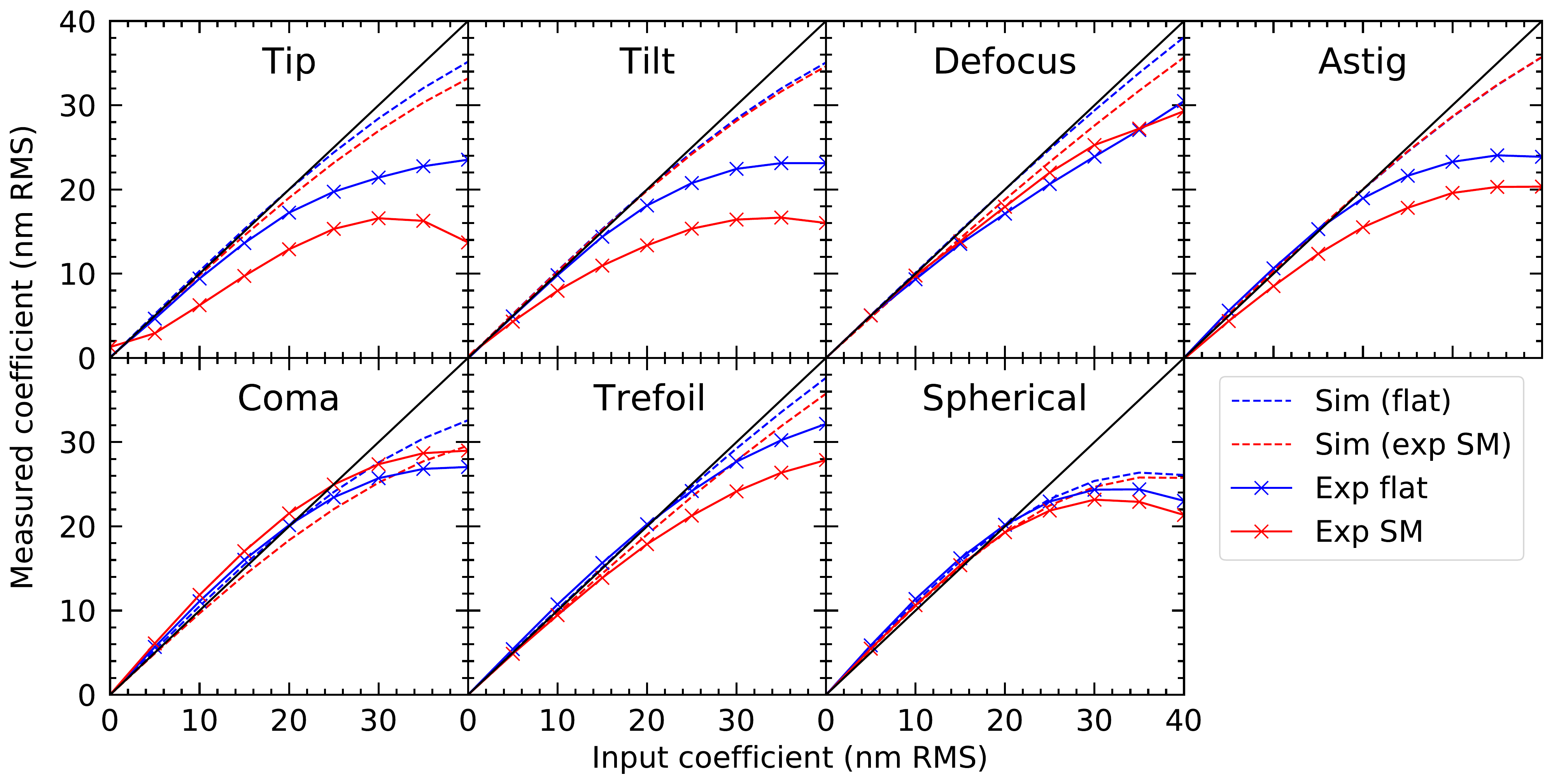}
    \caption{Response curves of the ZWFS to low-order Zernike modes, in simulation (dotted lines) and from HiCAT experiments (solid lines with cross markers). The ZWFS has been calibrated around the DM flat maps that optimize wavefront quality. Response curves are measured both around these flat maps (blue curves) and around SM commands (red curves).}
    \label{fig:response_curves}
\end{figure*}

In Fig.~\ref{fig:response_curves}, we present the ZWFS response curves for different commands with Zernike modes when operated with a Jacobian matrix to investigate the validity range of this data-driven calibration. These response curves are obtained by multiplying the intensity measurement on the ZWFS camera with the control matrix for different aberration amplitudes of the introduced probes. We recall that the calibration probes used to build the matrix $\mathbf{J}$ are performed using Zernike modes with $p$=20\,nm RMS.

In Fig.~\ref{fig:response_curves}, we compare the response curves to low-order Zernike modes in simulation and experiment for two base DM settings: the flat maps and the SM maps. The Zernike modes are generated in terms of requested OPD. Because of the linear approximation in the conversion from nm to volt and the unknown exact DM response, the exact introduced OPD is unknown. The ZWFS is calibrated with a Jacobian matrix built around the DM flat shapes. Except for tip and tilt aberrations, which we will discuss later on, the simulation and experiment results show good agreement for aberration amplitudes smaller than 15\,nm RMS. The slope of the curves is very close to the unity response. For larger aberrations, the experimental curves show a mode underestimation of typically 10 nm RMS for an input of 40 nm RMS. This effect most likely comes from the quadratic deflection-to-voltage relation of the DMs \citep{Morgan2019} that is not taken into account when applying DM commands on HiCAT. For open-loop operations like the ones shown here, we reach the limit of this DM control that assumes a perfect DM response. This approximation remains nevertheless sufficient for closed-loop control as detailed in Sec.~\ref{sec:closed_loop_operations}.

In the scope of future operations with changing DM offsets, an important point is to analyze the ZWFS behavior when we add an offset to the DM commands that were used during the Jacobian matrix calibration. For most modes, and as predicted by simulations, the sensor sensitivity is decreased when we perform the response measurement with the SM commands presented in Fig.~\ref{fig:sm_shapes}. Only the defocus and the coma show a slight increase. As for the tip and tilt modes, they present a larger loss in sensitivity even at small aberrations. This originates from the combination of two factors. First, the ZWFS robustness to these modes is known to be reduced compared to higher order modes, as they represent a  ZWFS mask misalignment with respect to the PSF. Second, the testbed drifts due to the delay between the SM command computation and the measurement of the response are likely to induce these tips and tilts. As a result, applying the SM commands that are computed around a slightly different testbed state will lead to a more visible effect on tip and tilt than on other modes.

For most science cases, the ZWFS will be operating in a closed loop with aberration values smaller that 20\,nm RMS. In this regime, the sensor shows a behavior that should not lead to the degradation of wavefront corrections, as its response remains linear. 

\subsection{Cross talk}

\begin{figure*}
    \centering
    \includegraphics[width=\linewidth]{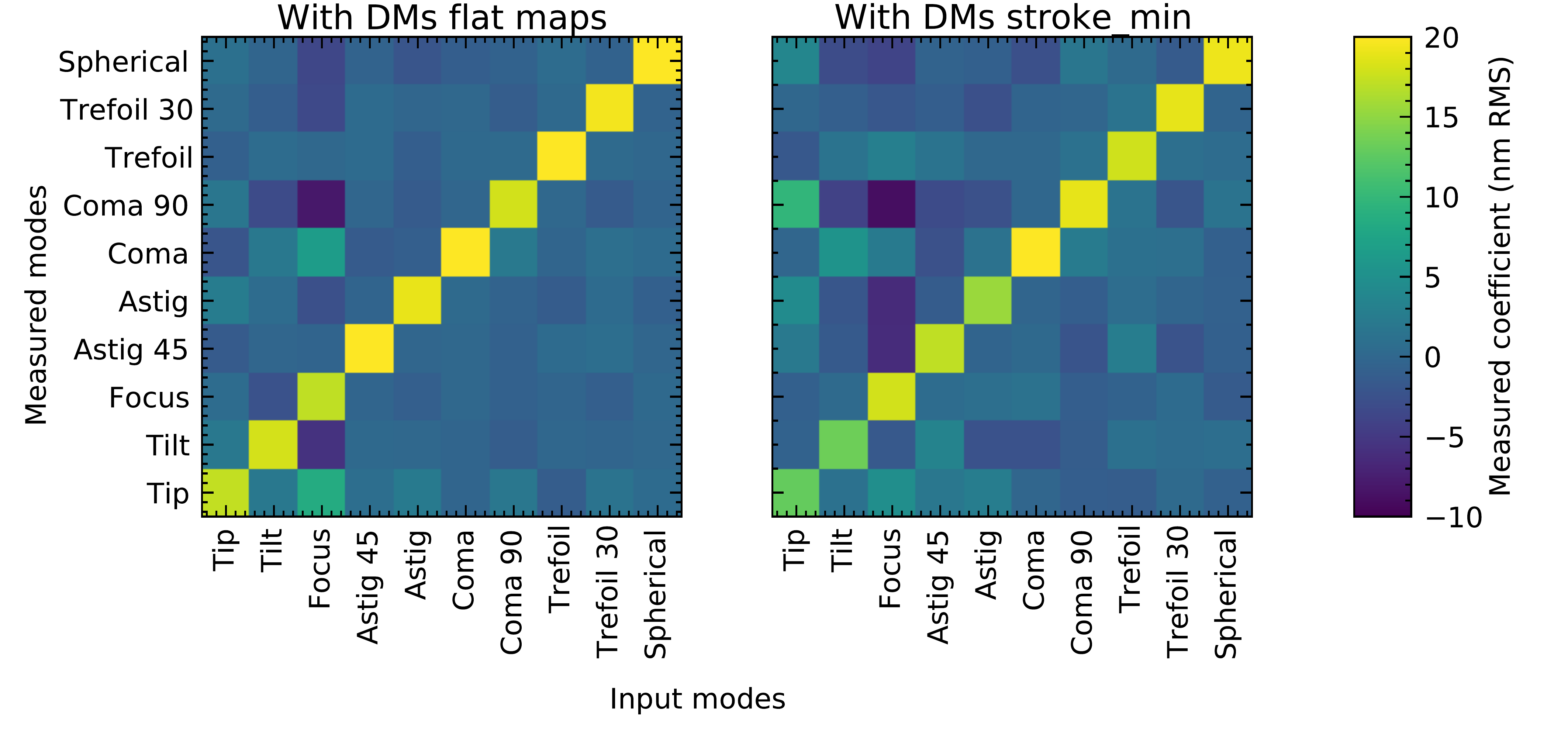}
    \caption{Mode coupling matrix of the ZWFS to Zernike modes using a Jacobian matrix calibrated with Zernike modes around flat DMs. Introduced and measured modes are given on the horizontal and vertical axes. Left: with flat DMs. Right: after introduction of SM offset on the DMs. Modes are introduced with 20 nm RMS in each case.}
    \label{fig:cross_talk_matrix}
\end{figure*}

On top of sensitivity, an important aspect of the sensor is cross-talk between sensed modes: do some modes trigger a response on other modes? Fig.~\ref{fig:cross_talk_matrix} shows the mode coupling matrix of the ZWFS for the first 10 measured Zernike modes. While we have $k=11$, only the first 10 modes are shown for the sake of simplicity. These measurements come from the same data set as Fig.~\ref{fig:response_curves} with $p$=20\,nm RMS. For both experiments, the global matrix shape is diagonal, which is consistent with the fact that the probes are identical to those used for the construction of $\mathbf{J}$. Outside the diagonal and for both experiment cases, defocus is measured as a combination of tip, tilt and both comas. We attribute this effect to a slight misalignment in the testbed that breaks the symmetry of the optical system, probably in the relay optics between the FPM and the ZWFS. For instance, a slight off-axis defocus could produce this effect. Since the non-linear response of the DMs to voltage is not taken into account, modal cross-talk can also be expected when adding an offset to the commands. Further tests are required to precisely identify the origin of this effect. 

The change of DM shape offsets keeps the coupling matrix similar between the two experiments and mainly diagonal. The main differences are created by the responses of the tip and tilt. These two modes show specifically a stronger coupling with both coma modes. The defocus also shows a coupling with both astigmatism modes that was not present with the offset of the flat maps. Overall, these limited changes due to DM offset modifications will unlikely impact the stability of the control loop, as we show in the following section.

\section{Stability results}
In this section, we investigate the wavefront and DH contrast stability with the ZWFS loop for different conditions of wavefront perturbations.

\subsection{WFE calibration and stability}

HiCAT is operating in air and naturally suffers from air turbulence, in particular in the optical path before the FPM. The air turbulence contribution is mainly composed of low-order aberrations, which allows us to test the ZWFS loop on an unknown perturbation and correct for them with DM1. With the air valve, we can adjust the strength of the dry air flow and open or close the testbed enclosure in order to generate different levels of air turbulence.

We study the wavefront statistics between an open-loop configuration with only WFE monitoring and a closed-loop configuration that includes WFE correction. We explore four different configurations, referring to the amount of turbulence. The first one, "closed", corresponds to the case with the air valve completely closed and no air flow, which causes the bench temperature and humidity to drift, but it reduces the turbulence to a minimum. The second one, "low", is related to a minimum air flow through the valve to stabilize the bench in humidity and temperature, which is the standard way of operation on HiCAT. The third one, "medium", leads to the case with a small opening between the HiCAT enclosure covers and an increased dry air flow. The last one, "high", corresponds to the case with the removal of one lateral panel of the enclosure and dry air flow at the maximum. In this configuration, the PSF typically jitters around values of $\pm 0.4\,\lsd$ at frequencies typically below 10\,Hz. Note that these medium and high turbulence modes are never used in normal operations for HiCAT, and only used here to increase the perturbation and study the performance of our closed-loop ZWFS in worse conditions than nominal. 

\begin{figure*}
    \centering
    \includegraphics[width=\linewidth]{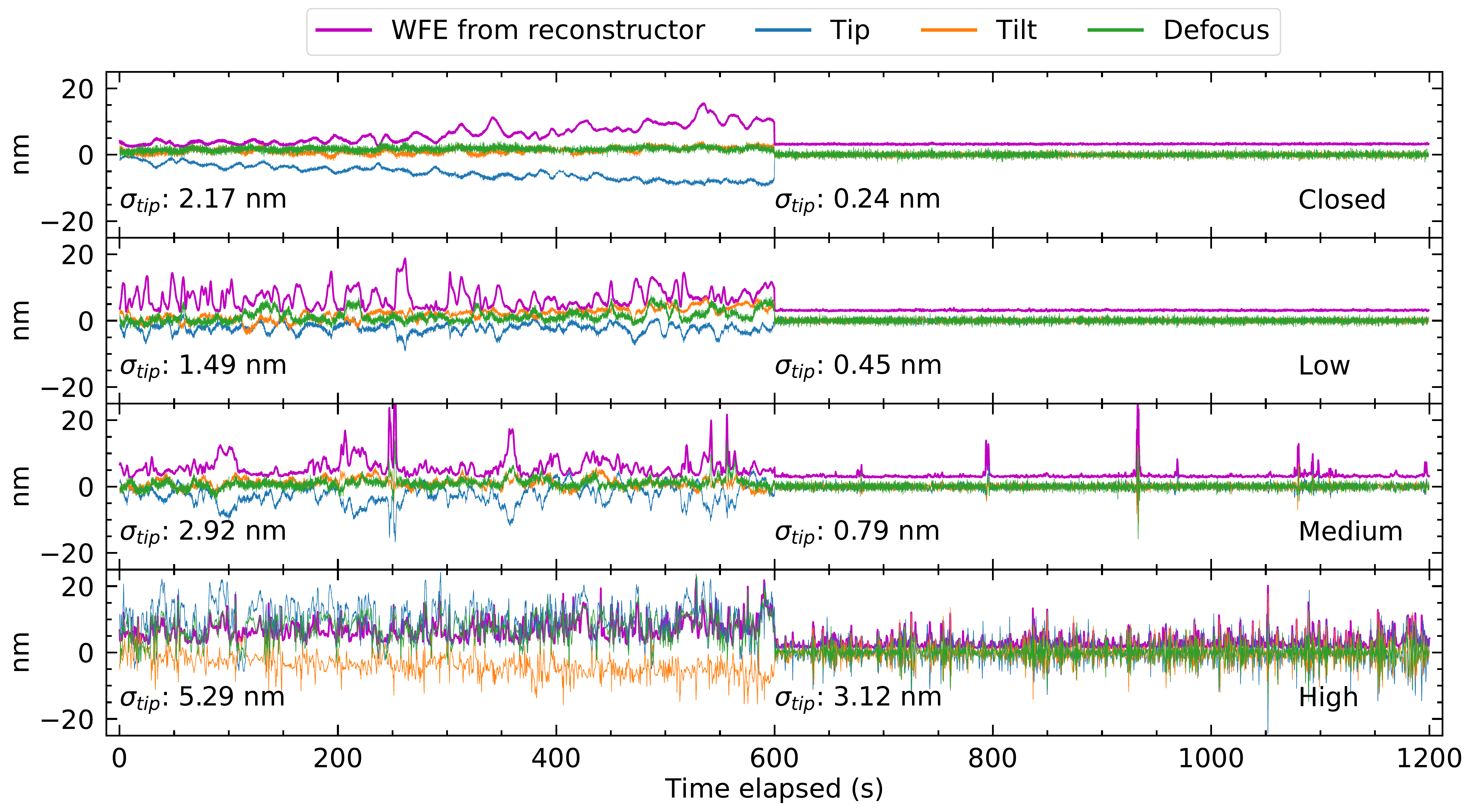}
    \caption{Temporal evolution of the wavefront error with different levels of air turbulence. Global WFE measured by 2nd-order reconstructor is plotted in magenta. Tip, tilt and defocus coefficients estimated by the inverse Jacobian matrix are plotted in blue, orange and green, respectively. The plots are ranked from top to bottom from no external air flow to maximum air flow. The experiment consisted of 15 minutes of open loop, followed by 15 minutes of closed loop, under natural drifting conditions. Nominal operations for HiCAT correspond to ``closed '' and ``low'', and the ``Medium/High'' configurations are only used to study the performance of the ZWFS outside of nominal regime.}
    \label{fig:air_turbulence}
\end{figure*}

In closed loop, the control is performed with a pre-built Jacobian matrix, with $m$=12 Zernike mode probes, $k$=11 modes being kept in the inversion. To avoid the effect of fitting error estimation on the whole WFE in the pupil as much as possible, the overall standard deviation of the wavefront is computed using the analytical phase reconstruction described in Sec.~\ref{sec:analytical_reconstruction}. Figure~\ref{fig:air_turbulence} shows the wavefront error estimates for the four different configurations of air turbulence on the testbed in open and closed loop. The closed loop runs at a typical frequency of 20$-$30\,Hz, limited by the current computer calculation speed. The differences of perturbation between all the situations are clearly visible in both open-loop, before 600 s, and in closed-loop, after 600 s.
For a closed air valve, a drift is present in open loop, degrading the tip mode, see top plot. This effect is due to a thermal drift of the testbed: the dry air is used to keep the temperature and humidity steady. With a restored air flow, this effect is not visible in the other configurations. The wavefront fluctuations become larger as the injected air turbulence increases.

Table~\ref{tab:turb_tab} gives the temporal standard deviation $\sigma$ for tip, tilt and defocus during the four experiments. From our experience on the testbed, tip is the most perturbed mode by air turbulence, but the behavior of other modes is very similar. The air turbulence degrades $\sigma_{tip}$ by a factor of 3.5, from 1.49 to 5.29\,nm RMS, between the "low" and "high" turbulence configurations. In all four configurations, closing the loop improves the residual tip, with the gain on $\sigma_{tip}$ ranging from 1.7 with the high turbulence, to 9 without any air flow. In the most stable configuration with no air flux, $\sigma_{tip}$ drops from 2.17 to 0.24\,nm RMS (and typically from $\pm0.2\,\lsd$ to $\pm0.05\,\lsd$), reaching sub-nanometric stability with our control loop. Improvements are also noticed with the other modes, confirming the ability of our control loop to correct for tip, tilt and defocus and enhance the stability of the first Zernike modes in all configurations. A similar behavior is also observed with higher order modes as illustrated in the following.

\begin{table}[]
    \centering
    \caption{Temporal WFE standard deviation in open and closed loop in nanometers, and the factor of improvement between both configurations for the tip, tilt and defocus modes and for each air turbulence configuration.}
    \begin{tabular}{ccccc}
        \hline
        \hline
        Valve & Aberration & Open loop & Closed loop & Gain  \\
        \hline
        \multirow{3}{*}{Closed}& Tip & 2.17 & 0.24 & 9.0\\
                                & Tilt & 0.91 & 0.48& 1.9\\
                                & Defocus & 0.70 & 0.64& 1.1\\ \hline
        \multirow{3}{*}{Low}& Tip & 1.49 & 0.45 & 3.3\\
                                & Tilt & 1.80 & 0.45& 4.0\\
                                & Defocus & 1.75 & 0.66& 2.6\\ \hline
        \multirow{3}{*}{Medium}& Tip & 2.92 & 0.79 & 3.7\\
                                & Tilt & 1.45 & 0.78& 1.9\\
                                & Defocus & 1.57 & 0.96& 1.6\\ \hline
        \multirow{3}{*}{High}& Tip & 5.29 & 3.11 & 1.7\\
                                & Tilt & 3.28 & 2.74& 1.2\\
                                & Defocus & 3.83 & 2.20& 1.7\\ \hline
    \end{tabular}
    \label{tab:turb_tab}
\end{table}

\begin{figure*}
    \centering
    \includegraphics[width=\linewidth]{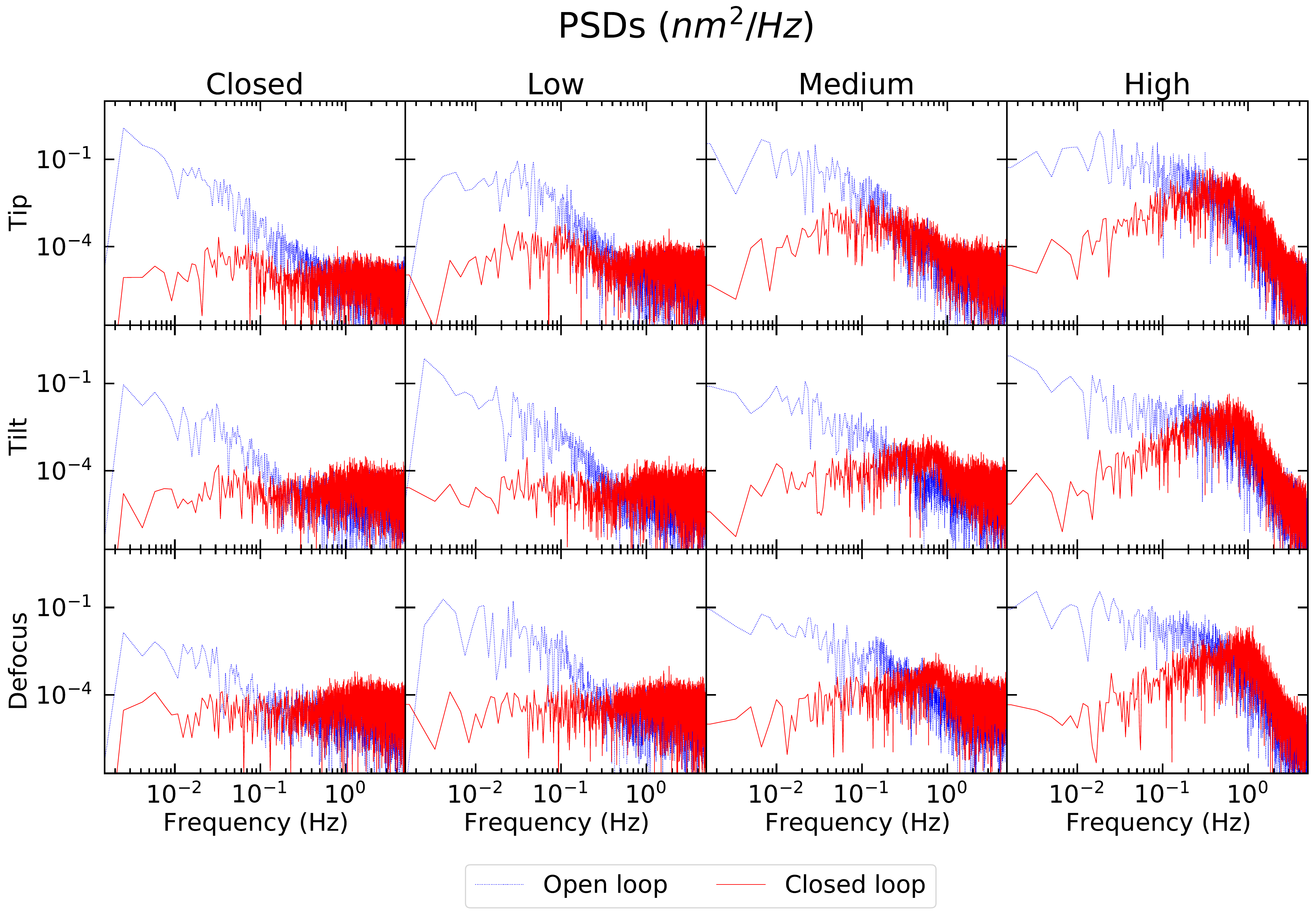}
    \caption{Temporal PSDs for the tip, tilt, and defocus in different air turbulence conditions, in open-loop (blue) and closed-loop (red) operations.}
    \label{fig:psds}
\end{figure*}

We also perform a temporal spectral analysis of our results, see Fig.~\ref{fig:psds}. The plot shows the open-loop (blue) and closed-loop (red) power spectrum densities (PSDs) as a function of the temporal frequency $f$ for the first three Zernike modes in the four different air turbulence regimes. All open-loop experiments show a characteristic decrease of the PSD that is consistent with the  $f^{-5/3}$ Kolmogorov statistics before reaching the noise floor of the higher temporal frequencies. The turbulence regime alters the frequency cutoff of this behavior, from around 0.2\,Hz for the closed valve case to several Hertz for the high turbulence regime. The PSDs in the open-loop operation do deviate from this $f^{-5/3}$ behavior in the low temporal frequency regime, most likely due to the bench enclosure that limits the turbulence developments at low spatial and temporal frequencies.

Still, all of the four closed-loop cases show an improvement of the temporal PSD in the low-frequency regime by two orders of magnitude from open to closed loop. For the ``closed" and ``low" air valve cases, the closed-loop PSDs are almost flat over all frequencies and reach the noise floor, meaning that the correction loop is close to the optimal correction in these conditions. The correction could be further improved by enhancing the signal-to-noise ratio in the measurements. This could be done by binning the images on the ZWFS camera, currently of size 224x224 pixels after 4x4 binning, or using temporal averages of several frames, at the cost of a lower loop speed, currently around 20$-$30 Hz. For medium and high turbulence regimes, the closed-loop PSDs show an increase from the lowest frequencies to 1\,Hz at which the blue and red curves meet, showing that the loop struggles to correct for the aberrations with frequencies higher than 0.1\,Hz. A larger gain, a faster loop speed or, an improved control method such as a proportional-integral-derivative (PID) controller or predictive control could be useful to improve the stability in these conditions. This high level of air turbulence is however never met in standard HiCAT operation. Overall, our control loop using ZWFS improves the temporal PSDs of the wavefront errors in all conditions, showing that it is a suitable solution for controlling and stabilizing low-order modes for high-contrast instruments with LCs.

\subsection{Contrast stability} \label{sec:dh_maintenance}

\begin{figure*}
    \centering
    \includegraphics[width=\linewidth]{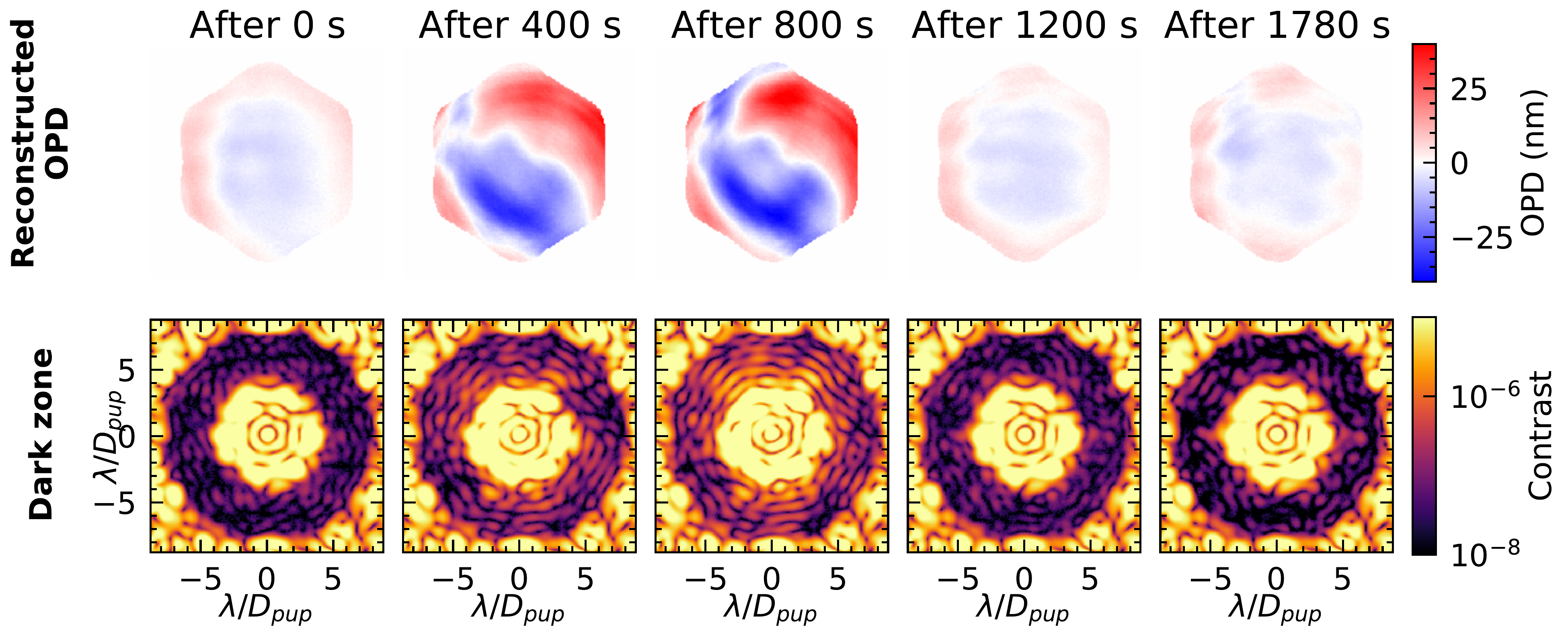}
    \caption{Examples of wavefront differences from reference as reconstructed by the ZWFS (top) and of DH images (bottom) during the contrast stabilization experiment with the calibration around flat maps. The iterations at 0\,s, 400\,s and 800\,s are from the open-loop experiment. The loop is closed after 900\,s. In this experiment we are adding ``artificial drifts'' in the form of low-order aberration random walk added to the continuous DM.  At the end of the open-loop phase (900\,s) the added aberration is about 20\,nm RMS. At the end of the closed-loop phase (1800\,s), the added aberration is about 25\,nm RMS. This experiment also corresponds to the nominal turbulence regime (``low'').}
    \label{fig:example_wf_dh}
\end{figure*}

The ultimate goal of correction for low-order aberrations is to help maintaining the optimal contrast in the DH. We explore the contrast evolution in a set-and-forget mode: we use the SM algorithm to compute optimal DM commands that allow for a DH at a few 10\up{-8} contrast, apply these commands on the DMs and let the DH contrast drift, with or without the ZWFS loop running. 
With the current large IWA ($4.5\,\lsd$), the DH contrast is dominated by mid- to high-order aberrations. The impact of the air turbulence and the ZWFS correction is therefore challenging to observe.
To artificially increase the contrast drift over the experiment duration of 30 min, we introduce Brownian noise on DM1. This random walk is performed every other second by drawing random Zernike coefficients with a uniform law in the range [-0.5, 0.5] nm RMS for the first 10 Zernike modes. The average WFE of each individual perturbation is 0.79\,nm RMS and the perturbations are introduced on DM1 in a cumulative way. We monitor the wavefront errors and the DH contrast first in open loop and then in close loop during 900\,s each. Figure~\ref{fig:example_wf_dh}, top plot, shows examples of wavefront maps measured with our ZWFS during the experiment. A clear degradation of the wavefront error can be observed during the open-loop experiment with a residual WFE of 3.7, 16.5 and 19.2\,nm RMS at 0, 400, and 800\,s respectively. After closing the loop at 900\,s, the wavefront errors show a pattern very similar to the initial one at 0\,s, with a WFE of 3.7\,nm RMS at 1780\,s, showing a good recovery of the initial wavefront state.

\begin{figure*}
    \centering
    \includegraphics[width=.9\linewidth]{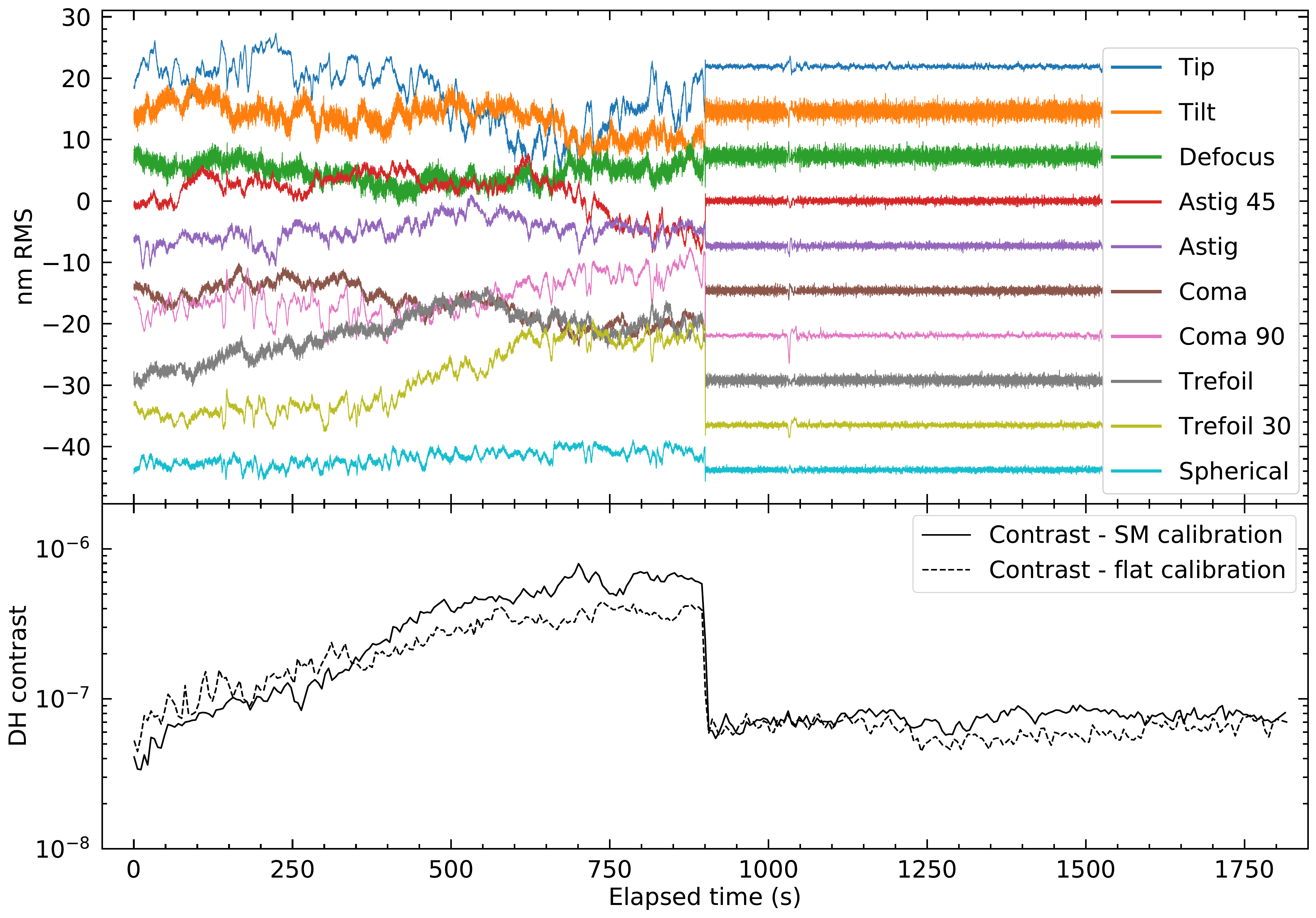}
    \caption{Wavefront evolution and contrast in the DH as a function of time in open-loop (first 900\,s) and in closed loop (last 900\,s) under artificial drift conditions. An additional random perturbation of 10 Zernike modes combined of 1.5\,nm RMS on average is applied every 2\,s. This amounts to a total aberration of about 25\,nm RMS at the end of the 30 minute experiment}. Top: Zernike coefficients estimated with the inverse Jacobian matrix calibrated around SM commands. They have been artificially shifted along the vertical axis to enhance readability: all of them actually oscillate around zero in closed loop. Bottom: Mean contrast in the DH in the case of a Jacobian matrix calibration performed around SM commands (continuous line) and around flat DM maps (dashed line). 
    \label{fig:contrast_stability}
\end{figure*}

The evolution of the corresponding first 10 Zernike coefficients estimated with the command matrix $\mathbf{C}$ are presented in Fig.~\ref{fig:contrast_stability}, top plot, for the calibration around SM commands. As expected, the coefficients show an improved stability after closing the loop. The temporal standard deviations are given for the first 10 modes in Table~\ref{tab:dh_stab_wf}. Once the loop is closed, the noisiest mode is tilt with $\sigma_{tilt}=760$\,pm, the mean value of $\sigma$ being 380\,pm between all the modes. Overall, the  wavefront error from the 10 Zernike coefficients drops from 10.4 to 1.4\,nm RMS, showing an improvement by a factor of 7.6. This improvement is confirmed by the analysis of the temporal PSDs for the first 10 Zernike coefficients, see Fig.~\ref{fig:dh_m_psd} in the appendix. A gain of a few orders of magnitude is observed in the low-temporal frequency regime from open to closed-loop operation.

\begin{table}[]
    \centering
    \caption{Open-loop and closed-loop temporal wavefront error standard deviations in nanometers for the first 10 Zernike modes controlled during the DH stabilization experiment, and the gain between both operations.}
    \begin{tabular}{cccc}
        \hline
        \hline
        Aberration & Open-loop & Closed-loop & Gain\\
        \hline
        Tip & 4.86 & 0.23& 20.92\\
        Tilt & 2.47 & 0.76& 3.24\\
        Defocus & 1.73 & 0.72& 2.40\\
        Astig 45 & 3.11 & 0.29& 10.60\\
        Astig & 1.92 & 0.28& 6.78\\
        Coma & 2.88 & 0.34& 8.60\\
        Coma 90 & 3.06 & 0.22& 13.66\\
        Trefoil & 3.64 & 0.43& 8.51\\
        Trefoil 30 & 5.20 & 0.22& 23.47\\
        Spherical & 1.21 & 0.24& 5.05\\
        \hline 
        RMS & 10.3 & 1.4 &7.6 \\
        \hline
    \end{tabular}
    \label{tab:dh_stab_wf}
\end{table}

\begin{table}[]
    \centering
    \caption{Statistics of the contrast spatial average in the DH, ranging from 4.5\,\lsd to 7.5\,\lsd, during the stabilization experiments for flat and DH calibrations.}
    \begin{tabular}{cccc}
        \hline
        \hline
        Calibration & Loop status & Mean & $\sigma_{contrast}$ \\
        \hline
        \multirow{2}{*}{DH}& Open & 3.3e-07 & 2.2e-07 \\
                           & Closed & 7.5e-08 & 7.4e-09 \\
        \hline
        \multirow{2}{*}{Flat}& Open & 2.4e-07 & 2.2e-07 \\
                             & Closed & 6.2e-08 & 8.2e-09 \\
        \hline
    \end{tabular}
    \label{tab:dh_stab_cont}
\end{table}

The bottom plot in Fig.~\ref{fig:example_wf_dh} displays focal-plane DH images corresponding to the wavefront measured by our ZWFS. The contrast degradation is visible at 400\,s and 800\,s, but the ZWFS closed loop restores a DH qualitatively similar to the initial one. The bottom plot in Fig.~\ref{fig:contrast_stability} shows the contrast evolution for two different calibrations, one where $\mathbf{J}$ was computed around the SM commands (solid line) and one where it was computed around flat commands (dashed line). The SM commands applied during both experiments however were the SM commands, at all times.

Regarding the contrast, we recall that our goal is to study the maintenance of a pre-computed DH with a given contrast and shape: the ZWFS control loop does not aim to improve the ultimate contrast or DH digging capabilities. As of December 2021, the capabilities of HiCAT allows for the generation of a DH with an average contrast of $\sim 2.5\times10^{-8}$ while SM is running. Once we interrupt the SM loop, due to the environmental testbed conditions and without any further action on the testbed, the contrast will tend to remain at twice this value, around $\sim 5\times10^{-8}$, over a few tens of minutes. As such, we are first and foremost limited by the contrast in the starting DH. After introducing some known perturbations that lead to a contrast degradation, we aim to recover the initial $\sim 5\times10^{-8}$ contrast thanks to our ZWFS-based control loop.

The spatially averaged contrast values in the DH are given in Table~\ref{tab:dh_stab_cont}. While in open loop, the contrast decreases by a factor of 10 from $5\times 10^{-8}$ to $5\times 10^{-7}$ due to the introduced wavefront drifts during the experiment; closing the loop brings the contrast back to approximately its initial value around $5-6\times10^{-8}$. In this experiment, the average closed-loop contrast for the ZWFS calibration around SM commands and flat maps are $7.6\times 10^{-8}$ and $6.2\times 10^{-8}$, respectively. We achieve a contrast standard deviation $\sigma_{contrast}$ that goes from $2.2 \times 10^{-7}$ in open loop to around $7\times10^{-9}$ in closed loop for both calibrations, which corresponds to an improvement by a factor of 30. We identify at least two reasons to explain the small difference between the initial ($5\times 10^{-8}$) and recovered ($6.2$ and $7.5\times 10^{-8}$) contrasts. First, as perturbations are still introduced during the closed-loop operation, the lag in the control loop will cause some residual WFE. While our loop is running at 20-30\,Hz, a faster loop will help reducing the lag and this residual effect. Second, while we correct for low-order modes, there are also mid- and high-order aberrations drifts that remain uncontrolled here, degrading the contrast DH with speckles. Although the spatially-filtered ZWFS approach will only correct for low-order aberrations, it can be used in conjunction with other high-order techniques to make the most efficient use of the rejected stellar photons. 

In Fig.~\ref{fig:contrast_stability} (bottom), the difference between the two curves is due to realization effects: the SM commands calculated at a given time do not provide optimal DH contrast at a different point in time. The achieved contrast also greatly depends on the bench evolution between the moment the SM commands are computed and the acquisition of the reference measurement for the ZWFS. This is clearly visible in Fig.~\ref{fig:example_wf_dh}: the initial DH, at 0\,s, shows some speckles that are usually removed by SM. Still, the contrast degradation at 800\,s due to the wavefront drift is canceled after the closed-loop control.  

Overall, we show that the stabilization of low-order aberrations on HiCAT is possible down to a sub-nanometric level and we demonstrate the efficiency of our control loop with ZWFS to prevent low-order WFE from degrading the contrast during observations. This correction has been performed in a blind way without interactions between the different control loops: our control loop based on the ZWFS uses no information from the imaging camera and it therefore stabilizes the image DH without seeing it. Further work involving synchronization with SM and running parallel loops on HiCAT will be investigated to stabilize the testbed at its best contrast performance.

\section{Conclusion and perspectives}

We have validated an approach to control low-order wavefront errors in Lyot-type coronagraphs by using a ZWFS with the light rejected by the coronagraphic FPM. In this work, the existing formalism of the ZWFS has been extended for wavefront reconstruction with a FPM-filtered beam.

To validate our approach experimentally, we have used the HiCAT testbed in such a configuration. We have characterized the ZWFS response to Zernike modes when calibrated with a modal Jacobian matrix. The ZWFS shows good response to low-order Zernike modes and limited cross-talk between modes. Changing the offsets on the DMs to generate a DH with a $2.5\times10^{-8}$ contrast and a 4.5\,\lsd IWA in the focal plane barely impacts the ZWFS response. With deeper contrasts and smaller IWAs, the DH DM commands may have larger strokes. Additional tests will then be necessary to extend the validity of our findings.

We validated the modal control of low-order aberrations on HiCAT by closing the loop between the ZWFS and DM1. In different air turbulence regimes, we show a clear wavefront correction and stabilization down to a temporal standard deviation of a few hundreds of picometers for the tip, tilt and defocus modes. In the presence of artificial random drifts introduced on DM1, our control loop is able to blindly stabilize the DH contrast at a level of $7\times10^{-8}$ with a standard deviation of $7\times10^{-9}$.

Further investigations are required to demonstrate a full operational capability. In this paper, only 11 Zernike modes were controlled. In the case of HiCAT, with a FPM mask cutoff radius of $4.2\lsd$, more Zernike modes could be measured. Regarding wavefront stability, controlling more modes is possible in general. However, these higher-order modes include a higher spatial frequency content that is filtered by the FPM and therefore not measurable by this ZWFS. Unconstrained by the control loop, these higher-order components add aberrations that create light leakage in the DH, which is not stable anymore. An improved modal control would have to be used with modes describing the system better as envisioned in \citet{Laginja2022}, or with Karhunen-Loeve modes. Furthermore, we currently use a monochromatic light source with enough photons for a high signal-to-noise ratio. A more realistic approach would be to validate the ZWFS in broadband light and in a photon-starved regime. 

The experiments were conducted on HiCAT with a segmented aperture that was flattened throughout with a calibration map. A crucial step will be to study the response of the ZWFS to segment piston, tip and tilt aberrations, and the correction in combination with the low-order modes. 

Finally, we worked in a set-and-forget setting for the DMs SM commands. The next step will consist of operating ZWFS while running SM and possibly other experiments in parallel. Our wavefront control loop currently runs at 30\,Hz limited by the camera readout speed and the computer processing time. The speed could be improved with a faster camera and software improvements.

This first demonstration of a control loop with ZWFS in the rejected light of a CLC and in the presence of a segmented aperture proves very promising to prepare for space coronagraphs to observe exo-Earth candidates in line with the recent recommendations of the 2020 Decadal Survey for a future segmented space observatory with high-contrast imaging and spectroscopy.

\begin{acknowledgements}
R.P. acknowledges PhD scholarship funding from R\'egion Provence-Alpes-C\^ote d'Azur and Thales Alenia Space.
The authors are especially thankful to the extended HiCAT team (over 50 people) who have worked over the past several years to develop this testbed. This work was supported in part by the National Aeronautics and Space Administration under Grant 80NSSC19K0120 issued through the Strategic Astrophysics Technology / Technology Demonstration for Exoplanet Missions Program (SAT-TDEM; PI: R. Soummer).  E.H.P. was supported by the NASA  Hubble Fellowship grant HST-HF2-51467.001-A awarded by the Space Telescope Science  Institute, which is operated by the Association of Universities for Research in Astronomy, Incorporated, under NASA contract NAS5-26555.

\end{acknowledgements}

\bibliographystyle{aa}
\bibliography{biblio}

\appendix

\section{Temporal PSD for the low-order modes during the DH maintenance experiment} \label{sec:appendix}

\begin{figure*}
    \centering
    \includegraphics[width=\linewidth]{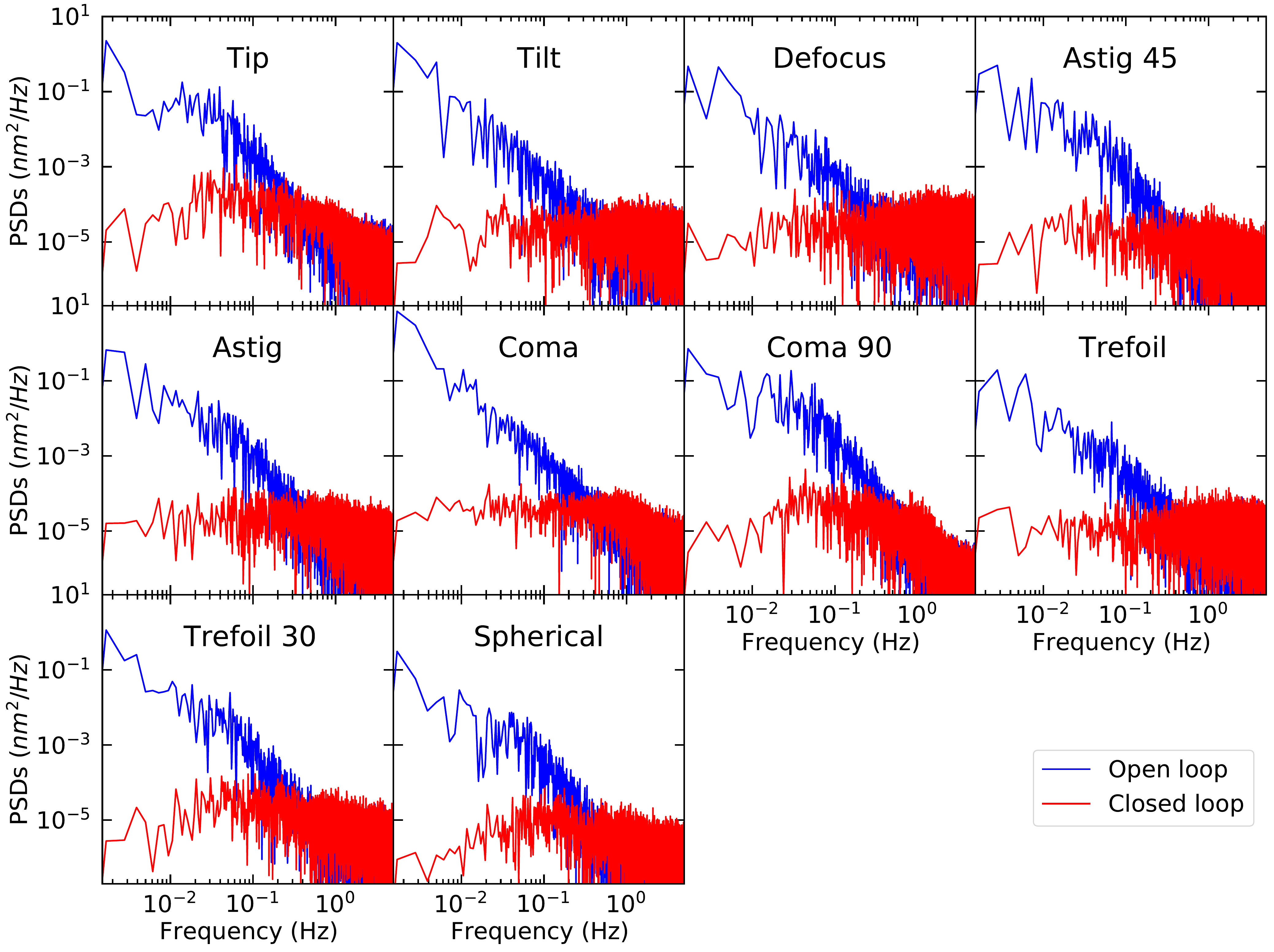}
    \caption{Temporal PSDs for 10 Zernike modes from tip to spherical aberration, in open-loop (blue) and closed-loop (red) operations. The data are related to the experiment presented in Sec.~\ref{sec:dh_maintenance}.} 
    \label{fig:dh_m_psd}
\end{figure*}

Fig.~\ref{fig:dh_m_psd} shows the temporal PSDs corresponding to the experiment presented in Sec.~\ref{sec:dh_maintenance}. The impact of the random walk is visible in the open-loop operation (blue curves) when comparing the PSDs of the tip, tilt and defocus in this plot and in Fig.~\ref{fig:psds} with the "low" air configuration. Observed at the lowest temporal frequencies, the PSD peaks reach $10^{-1}\,\mathrm{nm^2}/\mathrm{Hz}$ with the air turbulence only and about $1\,\mathrm{nm^2}/\mathrm{Hz}$ with the additional perturbations introduced on DM1. Overall, with the added perturbations, all the modes in Fig.~\ref{fig:dh_m_psd} present larger PSD values than the three modes shown in Fig.~\ref{fig:psds}, at all temporal frequencies, as expected.

In closed-loop operations, the final performance in Figs.~\ref{fig:dh_m_psd} and ~\ref{fig:psds} is similar with and without perturbations. Most modes present PSDs close to the noise floor around $10^{-5}\,\mathrm{nm^2}/\mathrm{Hz}$. The cutoff frequency of the control loop appears to be the same, at around 1\,Hz. As a result, Fig.~\ref{fig:dh_m_psd} shows a gain up to almost 5 orders of magnitude in the PSDs at low temporal frequencies from open to closed-loop operation. Tip, coma 90, trefoil 30 and spherical aberrations present a slight bump around $10^{-1}$\,Hz, possibly due to the fact that perturbations are introduced every other second. This analysis in the presence of perturbations validates the ability of our control loop with ZWFS to recover the aberration level that is observed in the absence of introduced perturbations.

\end{document}